\documentclass[11pt,a4paper]{article}
\usepackage{amsmath, color}
\usepackage{amssymb}
\usepackage{amsfonts}
\usepackage{amsthm}
\usepackage{bm}
\usepackage{mathrsfs}
\usepackage[all]{xy}
\usepackage{cite}
\usepackage{color}
\usepackage{graphicx}
\usepackage{braket}
\usepackage{authblk}
\usepackage[colorlinks,linkcolor=red,anchorcolor=blue,citecolor=blue]{hyperref}

\newcommand{\be}{\begin{equation}}
\newcommand{\ee}{\end{equation}}
\newcommand{\tr}{\mathrm{tr}}
\newtheorem{theorem}{Theorem}[section]
\newtheorem{lemma}[theorem]{Lemma}

\newtheorem{corollary}[theorem]{Corollary}

\theoremstyle{definition}

\theoremstyle{remark}

\numberwithin{equation}{section}

\begin{document}
\title{Distribution of spin correlation strengths in multipartite systems}
\author[1]{Bing Yu\thanks{mayubing59@mail.scut.edu.cn}}
\author[2]{Naihuan Jing\thanks{Corresponding Author: jing@ncsu.edu}}
\author[3]{Xianqing Li-Jost\thanks{Xianqing.Li-Jost@mis.mpg.de}}
\affil[1]{School of Mathematics, South China University of Technology, Guangzhou 510640, China}
\affil[2]{Department of Mathematics, North Carolina State University, Raleigh, NC 27695, USA}
\affil[3]{Max-Planck-Institute for Mathematics in the Sciences, 04103 Leipzig, Germany}
\renewcommand*{\Affilfont}{\small\it} 
\renewcommand\Authands{ and } 
\date{}
\maketitle
\begin{abstract}
For a two-qubit state the isotropic strength measures the degree of isotropic spin correlation.
The concept of isotropic strength is generalized to multipartite qudit systems, and the strength distributions
for tripartite and quadripartite qudit systems are thoroughly investigated.
We show that the sum of relative isotropic strengths of any three qudit state over $d$-dimensional Hilbert space cannot exceed $d-1$, which generalizes of the case $d=2$. The trade-off relations and monogamy-like relations of the sum of spin correlation strengths for pure three- and four-partite systems are derived. Moreover, the bounds of spin correlation strengths among different subsystems of a quadripartite state are used to analyze quantum entanglement.
\end{abstract}

\section{Introduction}

Let $\rho$ be a two-qubit state on
$\mathcal {H}_2^A\otimes\mathcal {H}_2^B$ in the Bloch form \cite{G}
\begin{equation}
\rho_{AB}=\frac1{4} (I\otimes I+\sum_{i=1}^{3}a_i{\sigma_i}\otimes I+\sum_{j=1}^{3}b_j I\otimes {\sigma_j}+\sum_{i,j=1}^{3}R_{ij}{\sigma_i}\otimes{\sigma_j}),
\end{equation}
where $\sigma_1,\sigma_2,\sigma_3$ are the Pauli spin matrices, $a_i=\tr(\rho_{AB}{\sigma_i}\otimes I)$, $b_j=\tr(\rho_{AB}I\otimes {\sigma_j})$, and $R_{ij}=\tr(\rho_{AB}{\sigma_i}\otimes{\sigma_j})$.
The {\it spin correlation matrix} $R=(R_{ij})$ and the vectors $\bm{a}=(a_1, a_2, a_3)^t$ and $\bm{b}=(b_1, b_2, b_3)^t$
characterize the two-qubit state in an essential way.
The three quantities are closely related to the intensity of quantum correlations \cite{H09,BCP,MBC}, and they are also
utilized in several fundamental concepts such as quantum entanglement
\cite{NC,CKW,BBL,V,HJ,GT,VH,MC,CMC,LJ,LS}, quantum discord\cite{OZ,DVB,DVC,LF}, EPR steering\cite{WJD,JPJ,MJJW,JHAZW,CM,ZCL} and Bell nonlocality \cite{B,CHSH,HHH,WQ}.

For multi-partite quantum state $\rho$, the spin correlation matrices among any two partite substates also reveal intrinsic properties
of the quantum phenomena.
For three-qubit pure state $\rho_{ABC}\in\mathcal {H}_2^A\otimes\mathcal {H}_2^B\otimes\mathcal {H}_2^C$, its three reduced two-qubit states $\rho_{AB}$, $\rho_{BC}$ and $\rho_{AC}$, and the associated spin correlation matrices $R^{AB}$, $R^{BC}$ and $R^{AC}$ have figured prominently in the recent interesting work of Cheng and Hall \cite{CH}. Therein they introduced the isotropic strength $s_{iso}^{AB}$
as the average of the three eigenvalues of the matrix $R^{AB}{R^{AB}}^t$ and showed that the sum of the three isotropic strengths $s_{iso}^{AB}$, $s_{iso}^{BC}$ and $s_{iso}^{AC}$ satisfies the amazing identity $s_{iso}^{AB}+s_{iso}^{BC}+s_{iso}^{AC}=1$, which can be used to deduce the volume monogamy relation of quantum steering ellipsoids \cite{CM} and strong monogamy relations for Bell nonlocality \cite{CH}.

In this work, we generalize the spin correlation matrix of the two-qubit systems to the general qudit system in order to reveal fundamental properties of the quantum system. Moreover, we would like to understand how variance of the Hilbert space affects the
situation, in the hope to learn the fundamental local unitary invariance. We generalize the isotropic strength from the three-qubit system to the tripartite and quadripartite qudit systems, and investigate the distributions of spin correlation strengths. For pure tripartite qudit systems $\mathcal {H}_d^A\otimes\mathcal {H}_d^B\otimes \mathcal {H}_d^C$, based on the purity of the reduced state of the pure three-qudit state, we obtain that the sum of the isotropic strengths for arbitrary three-qudit state over $d$-dimensional Hilbert space cannot exceed $d-1$, which is a generalization of one main identity in \cite{CH}. For pure quadripartite qudit systems, we give the trade-off relations, monogamy relations of the spin correlation strengths similarly in the tripartite case.

For a multipartite system our bounds of the spin correlation strengths among different subsystems can be utilized to analyze the intrinsic quantum entanglement. We first give necessary conditions of a pure four-qudit state being biseparable by using \eqref{Sisobound} and Corollary 2.2 (cf. \eqref{e:sum1}). After that
we generalize Vicente-Huber's method \cite{VH} of detecting genuine multipartite entanglement (GME) to all quantum four-qudit states.

This paper is organized as follows. In Sec. \ref{sec:leve2} we generalize the isotropic strength of the pure three-qubit systems to the tripartite qudit systems, and show the sum of isotropic strengths can not exceed $d-1$. We obtain the trade-off relation and other interesting properties of isotropic strengths. In Sec. \ref{sec:leve3} we extend the results to pure quadripartite qudit systems and present the trade-off relation for the spin correlation strengths.
In Sec. \ref{sec:leve4}, we show how the distribution of spin correlation strengths are used to detect quantum entanglement. Conclusion and summary are given in Sec. \ref{sec:leve5}.

\section{\label{sec:leve2}Distribution of spin correlation strengths for tripartite state}
Let $\rho_{AB}$ be the density matrix of a bipartite state on the tensor product  $\mathcal {H}_d^A\otimes\mathcal {H}_d^B$,
where $\mathcal {H}_d$ is an $d$-dimensional Hilbert space.
Let $\lambda_i$ be the Gell-Mann basis elements (self-dual) on $\mathcal H_d$
normalized as
$\tr(\lambda_i\lambda_j)=d\delta_{ij}$ and $\lambda_0=I$. Denote $\bm{\lambda}=(\lambda_1, \ldots, \lambda_{d^2-1})$. Then $\rho_{AB}$ can be written in the Bloch form
\begin{equation}\label{e:bl2}
\rho_{AB}=\frac1{d^2} (I\otimes I+\sum_{i=1}^{d^2-1}a_i{\lambda_i}\otimes I+\sum_{j=1}^{d^2-1}b_j I\otimes {\lambda_j}+\sum_{i,j=1}^{d^2-1}R_{ij}{\lambda_i}\otimes{\lambda_j}),
\end{equation}
where $a_i=\tr(\rho_{AB}{\lambda_i}\otimes I)$, $b_j=\tr(\rho_{AB}I\otimes {\lambda_j})$, and $R_{ij}=\tr(\rho_{AB}{\lambda_i}\otimes{\lambda_j})$. The vectors
\begin{equation}
\bm{a}=(a_1,a_2,\ldots,a_{d^2-1})^t, \bm{b}=(b_1,b_2,\ldots,b_{d^2-1})^t 
\end{equation}
are the Bloch vectors of the reduced states $\rho_A$ and $\rho_B$ respectively:
\begin{equation}\label{e:onepart}
\rho_A=\frac1{d}(I+\bm{a}\cdot\bm{\lambda})=\frac1{d}(I+\sum_{i=1}^{d^2-1}a_i {\lambda}_i).
\end{equation}
The matrix
$R=R^{AB}=(R_{ij})_{(d^2-1)\times (d^2-1)}$ is called the {\it spin correlation matrix} of $\rho_{AB}$. 

Clearly $\bm{a}, \bm{b}$ and $R^{AB}$ are
local unitary invariants of the state $\rho^{AB}$. In particular, if $\rho^{AB}$ is pure,
then $\mathrm{tr}\rho_{A}^2=\mathrm{tr}\rho_B^2$ implies that $\bm{a}^2=\bm{b}^2$. Also $\bm{a}^2\leqslant d-1$ due to the fact that $\mathrm{tr}\rho_{A}^2\leqslant1$.

The spin correlation matrix $R^{AB}$ reveals some of the characteristic properties of $\rho^{AB}$.
Consider the eigenvalues of $R^{AB}{R^{AB}}^t$ arranged in descending order: $s_1\geqslant s_2\geqslant \ldots \geqslant s_{d^2-1}$.
Generalizing the qubit case \cite{CH}, we define the {\it isotropic strength} of the density
matrix $\rho_{AB}$ as the average of the eigenvalues:
\begin{equation}
s_{iso}^{AB}=\frac1{d^2-1}\sum_{k=1}^{d^2-1}s_k=\frac{||R^{AB}||^2}{d^2-1},
\end{equation}
where $||R||=\sqrt{\mathrm{tr}(RR^t)}$ is the Frobenius norm.
\

Now we consider a general pure tripartite state $\rho_{ABC}=\ket{\psi}\bra{\psi}$ on $\mathcal {H}_d^A\otimes\mathcal {H}_d^B\otimes \mathcal {H}_d^C$, where $\langle \psi|\psi\rangle=1$. Its Bloch form relative to tensor products of the Gell-Mann basis $\lambda_i$ is
\begin{align}\label{e:tribloch}
\rho_{ABC}=\frac1{d^3}(&I\otimes I\otimes I+\sum_{i=1}^{d-1}a_i{\lambda_i}\otimes I\otimes I+\sum_{j=1}^{d-1}b_jI\otimes  \lambda_j\otimes I+\sum_{l=1}^{d-1}c_lI\otimes I\otimes
\lambda_l\notag\\
+&\sum_{i,j=1}^{d^2-1}R_{ij}^{AB}{\lambda_i}\otimes{\lambda_j}\otimes I+\sum_{j,l=1}^{d^2-1}R_{jl}^{BC}I\otimes {\lambda_j}\otimes{\lambda_l}+\sum_{i,l=1}^{d^2-1}R_{il}^{AC}{\lambda_i}\otimes I\otimes{\lambda_l} \notag\\
+&\sum_{i,j,l=1}^{d^2-1}R_{ijl}^{ABC}{\lambda_i}\otimes{\lambda_j}\otimes{\lambda_l}),
\end{align}
where the vectors $\bm a, R^{AB}, R^{ABC}$ etc. are taken as column vectors with the indices arranged in the lexicographic order.
Each entry of the component vector is given by trace function, for example, $R_{ijl}^{ABC}=\mathrm{tr}(\rho_{ABC}{\lambda_i}\otimes{\lambda_j}\otimes{\lambda_l}), i,j,l=1,\ldots, d^2-1$.
For uniformity the vectors $\bm a, \bm b, \bm c$ are also denoted by $R^A, R^B, R^C$ respectively.

The two-partite reduced states
are $\rho_{AB}=\tr_C\rho_{ABC}$ etc. (see \eqref{e:bl2}).
So $R^{AB}$, $R^{BC}$ and $R^{AC}$ still denote the spin correlation matrices respectively.
It follows from the Schmidt decomposition that the purity of tripartite state implies that
any bipartition of the pure state $\rho_{ABC}$ satisfies $\mathrm{tr}(\rho_{AB}^2)=\mathrm{tr}(\rho_C^2)$, $\mathrm{tr}(\rho_{BC}^2)=\mathrm{tr}(\rho_A^2)$ and $\mathrm{tr}(\rho_{AC}^2)=\mathrm{tr}(\rho_B^2)$.

\
Invoking the purity of any bipartition of the pure three qudit state $\rho_{ABC}$, we calculate the isotropic strengths as follows.
\begin{align}\label{Siso}
s_{iso}^{AB}=\frac{d(1+\bm c^2)-1-\bm a^2-\bm b^2}{d^2-1},\notag\\
s_{iso}^{BC}=\frac{d(1+\bm a^2)-1-\bm b^2-\bm c^2}{d^2-1},\\
s_{iso}^{AC}=\frac{d(1+\bm b^2)-1-\bm a^2-\bm c^2}{d^2-1}.\notag
\end{align}
Note that
\begin{equation}\label{Sisobound}
\|R^{AB}\|^2=s_{iso}^{AB}(d^2-1)\leqslant d-1+d\bm c^2\leqslant d^2-1,
\end{equation}
where the second inequality comes from $\bm c^2\leqslant d-1$,
similarly the inequality holds for $\|R^{AC}\|^2$ and $\|R^{BC}\|^2$ as well.

Subsequently for a pure tripartite state
\begin{equation}\label{e:isosum}
s_{iso}^{AB}+s_{iso}^{BC}+s_{iso}^{AC}=\frac{3d-3+(d-2)(\bm a^2+\bm b^2+\bm c^2)}{d^2-1},
\end{equation}

\begin{lemma}\label{le:1}
For a pure tripartite qudit state $\rho_{ABC}$
the invariants satisfy the following relation:
\begin{equation}\label{inv}
\bm a^2+\bm b^2+\bm c^2+\frac{\|R^{ABC}\|^2}{d-1}=(d+2)(d-1)
\end{equation}
where $\bm a$, $\bm b$, $\bm c$ are the Bloch vectors of the reduced states $\rho_A$, $\rho_B$, and $\rho_C$ respectively
and $\|R^{ABC}\|$ is the Euclidean norm.
\end{lemma}

\emph{Proof}. For a pure tripartite state $\rho_{ABC}$ given as in
\eqref{e:tribloch},
$\mathrm{tr}\rho_{ABC}^2=1$. Then $1+\bm a^2+\bm b^2+\bm c^2+(d^2-1)(s_{iso}^{AB}+s_{iso}^{BC}+s_{iso}^{AC})+\|R^{ABC}\|^2=d^3$.
Using \eqref{e:isosum} we see that the invariants satisfy the relation  (\ref{inv}). \qed

\begin{corollary}\label{c:2} For any tripartite state $\rho$, one has the following bound:
\begin{equation}\label{e:sum1}
\|R^{ABC}\|^2\leqslant (d-1)^2(d+2).
\end{equation}
\end{corollary}

In fact,  Lemma \ref{le:1} says that $\|R^{ABC}\|\leqslant (d-1)\sqrt{d+2}$ for a pure state $\rho$. Then for a general state
$\rho=\sum_\alpha p_\alpha\rho_\alpha$ with $\sum_\alpha p_\alpha=1$, the convex property of the Euclidean norm implies that
\begin{align}\notag
\|R^{ABC}\|&=\|\sum_\alpha p_\alpha R^{ABC}(\rho_\alpha)\|\leqslant \sum_\alpha p_\alpha\|R^{ABC}(\rho_\alpha)\|\\
&\leqslant\sum_\alpha p_\alpha (d-1)\sqrt{d+2}=(d-1)\sqrt{d+2}.
\end{align}

Using Lemma \ref{le:1}, we can derive the upper bound of the sum of the isotropic strengths of the pure three qudit state $\rho_{ABC}$  and the trade-off relation about isotropic strengths immediately.

\begin{theorem}\label{t:2}
For a pure tripartite qudit state $\rho_{ABC}$,
the sum of isotropic strengths has the following trade-off relation:
\begin{equation}\label{e:t2}
s_{iso}^{AB}+s_{iso}^{AC}+s_{iso}^{BC}=d-1-\frac{d-2}{(d+1)(d-1)^2}\|R^{ABC}\|^2.
\end{equation}
Thus $\|R^{ABC}\|^2$ can be viewed as a measure of the tripartite spin correlation strength.
\end{theorem}

Note that when $d=2$, \eqref{e:t2} reduces to $s_{iso}^{AB}+s_{iso}^{AC}+s_{iso}^{BC}=1$, which is one of the key relations discovered in
\cite{CH} for pure three-qubit state.

From  (\ref{Siso}) it follows that
\begin{equation}
\frac{1}{d-1}\bm a^2-s_{iso}^{BC}=\frac{1}{d-1}\bm b^2-s_{iso}^{AC}=\frac{1}{d-1}\bm c^2-s_{iso}^{AB}.
\end{equation}

We remark that a pure tripartite state has the following bounds:
$\frac3{d+1}\leq s_{iso}^{AB}+s_{iso}^{AC}+s_{iso}^{BC}\leq d-1$.

%

\begin{theorem}\label{t:1}
Let $\rho_{ABC}$ be a general tripartite quantum state
and $s_{iso}^{AB}$, $s_{iso}^{AC}$, $s_{iso}^{BC}$ the relative isotropic strengths of the reduced
states.
One has that $s_{iso}^{AB}+s_{iso}^{AC}+s_{iso}^{BC}\leqslant d-1$.
\end{theorem}

\emph{Proof}. It suffices to consider the pure state.
The upper bound is clear from Theorem \ref{t:2}.
\qed

Theorems \ref{t:2}-\ref{t:1} generalize the trade-off relations from three qubits to general quantum tripartite systems.

\section{\label{sec:leve3}Distribution of spin correlation strengths for quadripartite state}
In this section, we generalize the trade-off relations of spin correlation strengths to
any quadripartite quantum
state on $\mathcal {H}_d^A\otimes\mathcal {H}_d^B\otimes \mathcal {H}_d^C\otimes \mathcal {H}_d^D$.

Let $\rho_{ABCD}$ be a pure four-qudit state on $\mathcal{H}_d^{\otimes 4}$ in the Bloch form similarly
to \eqref{e:tribloch}, then its
reduced states $\rho_{ABC}$ and $\rho_D$ can be written respectively as in
 \eqref{e:tribloch} and  \eqref{e:onepart}. It follows from the purities of the bipartition reduced states that
$\mathrm{tr}(\rho_{AB}^2)=\mathrm{tr}(\rho_{CD}^2)$, $\mathrm{tr}(\rho_{AC}^2)=\mathrm{tr}(\rho_{BD}^2)$ and $\mathrm{tr}(\rho_{AD}^2)=\mathrm{tr}(\rho_{BC}^2)$. The following relations are then
easily seen for
the isotropic strengths of bipartitions:
\begin{align}\label{s4:1}
s_{iso}^{AB}-s_{iso}^{CD}=\frac{-(\bm a^2+\bm b^2)+\bm c^2+\bm d^2}{d^2-1},\notag \\
s_{iso}^{AC}-s_{iso}^{BD}=\frac{-(\bm a^2+\bm c^2)+\bm b^2+\bm d^2}{d^2-1},\\
s_{iso}^{AD}-s_{iso}^{BC}=\frac{-(\bm a^2+\bm d^2)+\bm b^2+\bm c^2}{d^2-1}.\notag
\end{align}
Similarly for the bipartitions $(ABC,D)$, $(ACD,B)$, $(BCD,A)$ and $(ABD,C)$, one also
has identical purities for any pair of the reduced states $\{\rho_{ABC}, \rho_{D}\}$, $\{\rho_{ACD}, \rho_{B}\}$, $\{\rho_{BCD}, \rho_{A}\}$, and $\{\rho_{ABD}, \rho_{C}\}$. Subsequently we have that
\begin{align}\label{s4:2}
s_{iso}^{BC}+s_{iso}^{AC}+s_{iso}^{AB}=\frac{d^2(1+\bm d^2)-(1+\bm a^2+\bm b^2+\bm c^2)-\|R^{ABC}\|^2}{d^2-1},\notag\\
s_{iso}^{CD}+s_{iso}^{AC}+s_{iso}^{AD}=\frac{d^2(1+\bm b^2)-(1+\bm a^2+\bm c^2+\bm d^2)-\|R^{ACD}\|^2}{d^2-1},\\
s_{iso}^{BC}+s_{iso}^{CD}+s_{iso}^{BD}=\frac{d^2(1+\bm a^2)-(1+\bm b^2+\bm c^2+\bm d^2)-\|R^{BCD}\|^2}{d^2-1},\notag\\
s_{iso}^{AB}+s_{iso}^{AD}+s_{iso}^{BD}=\frac{d^2(1+\bm c^2)-(1+\bm a^2+\bm b^2+\bm d^2)-\|R^{ABD}\|^2}{d^2-1},\notag
\end{align}
where $\|R^{ABC}\|^2$ is the tripartite spin correlation strength of the reduced state $\rho_{ABC}=\mathrm{\mathrm{tr}}_D(\rho_{ABCD})$
and the other spin correlation strengths $\|R^{ACD}\|^2$, $\|R^{BCD}\|^2$, or $\|R^{ABD}\|^2$ are defined similarly.

\

Simple calculation leads to the relation between the four tripartite spin correlation strengths, for example,
\begin{align} \nonumber
&(d^2-1)\bm a^2-\|R^{BCD}\|^2=(d^2-1)\bm b^2-\|R^{ACD}\|^2\\  \label{eq3.3}
&=(d^2-1)\bm c^2-\|R^{ABD}\|^2=(d^2-1)\bm d^2-\|R^{ABC}\|^2.
\end{align}
Therefore we have that
\begin{align}
\|R^{ACD}\|^2-\|R^{BCD}\|^2&=(d^2-1)(\bm b^2-\bm a^2),\\
\|R^{ABC}\|^2-\|R^{ABD}\|^2&=(d^2-1)(\bm d^2-\bm c^2),\\
\|R^{ABD}\|^2-\|R^{ACD}\|^2&=(d^2-1)(\bm c^2-\bm b^2).
\end{align}

\begin{theorem}\label{t:3}
For a pure quadripartite state $\rho_{ABCD}$ over $\mathcal {H}_d^A\otimes\mathcal {H}_d^B\otimes \mathcal {H}_d^C\otimes \mathcal {H}_d^D$, the isotropic strengths of the state satisfy the following trade-off relation,
\begin{align} \nonumber
&s_{iso}^{AB}+s_{iso}^{AC}+s_{iso}^{AD}+s_{iso}^{BC}+s_{iso}^{CD}+s_{iso}^{BD}\\
&=\frac{(-d^2+3)(d^2-1)+(d^2-2)(\bm a^2+\bm b^2+\bm c^2+\bm d^2)+\|R^{ABCD}\|^2}{d^2-1},
\end{align}
where $\|R^{ABCD}\|$ is the Euclidean norm of the vector $R^{ABCD}$.
\end{theorem}

As $\|R^{ABCD}\|^2$ is a measure of the quadripartite spin correlation strength, the trade-off relation implies that
the quadripartite spin correlations are tied up with
relative isotropic strengths of the four-partite qudit state $\rho_{ABCD}$.

\

\begin{corollary}\label{C:1}
For a pure quadripartite state $\rho_{ABCD}$,
the sum of isotropic strengths also satisfies the trade-off relation,
\begin{align}
&s_{iso}^{AB}+s_{iso}^{AC}+s_{iso}^{AD}+s_{iso}^{BC}+s_{iso}^{CD}+s_{iso}^{BD}\\
=&\frac{2(d^2-1)-(\bm a^2+\bm b^2+\bm c^2+\bm d^2)-2((d^2-1)\bm a^2-\|R^{BCD}\|^2)}{d^2-1}\notag.
\end{align}
\end{corollary}
Proof. The above trade-off relation can be obtained easily by combining  \eqref{s4:2} with  \eqref{eq3.3}.

\begin{corollary}\label{C:2}
For any pure four-qudit state $\rho_{ABCD}$,
the sum of isotropic strengths also satisfies the monogamy relation:
\begin{equation}\label{monog}
s_{iso}^{AB}+s_{iso}^{AC}+s_{iso}^{AD}=\frac{d^2-1+(d^2-3)\bm a^2-\|R^{BCD}\|^2}{d^2-1}.
\end{equation}
\end{corollary}

Remark. Similar monogamy relations are easily obtained when taking one of other particles $B$, $C$ and $D$ as a central one.
For example, when particle $B$ is treated as central, the following monogamy relation follows immediately,
\begin{equation}
s_{iso}^{BA}+s_{iso}^{BC}+s_{iso}^{BD}=\frac{d^2-1+(d^2-3)\bm b^2-\|R^{ACD}\|^2}{d^2-1}.
\end{equation}

Note that for $d=2$ it follows from Theorem \ref{t:1} that each equality in  \eqref{s4:2} cannot exceed 1, sum of isotropic strengths for the four-qubit pure state satisfies the trade-off relation:
\begin{align}\label{4qub}
&s_{iso}^{AB}+s_{iso}^{AC}+s_{iso}^{AD}+s_{iso}^{BC}+s_{iso}^{CD}+s_{iso}^{BD}\notag\\
=&\frac{12+\bm a^2-\|R^{BCD}\|^2+\bm b^2-\|R^{ACD}\|^2+\bm c^2-\|R^{ABD}\|^2+\bm d^2-\|R^{ABC}\|^2}{6}\leqslant2.
\end{align}
In this case, one also has that
\begin{equation}
\bm a^2+\bm b^2+\bm c^2+\bm d^2\leq \|R^{BCD}\|^2+\|R^{ACD}\|^2+\|R^{ABD}\|^2+\|R^{ABC}\|^2.
\end{equation}

\

\section{\label{sec:leve4} Quadripartite entanglement detection}

In this section, we discuss how our bounds of spin correlation strengths among different subsystems are 
applied in detecting entanglement for quadripartite quantum states.
We consider the quadripartite space $\otimes_{k=1}^4\mathcal{H}_k$, where $\mathcal H_k\simeq \mathbb C^d$ is the space for the $k$th particle.

A density matrix $\rho$ on $\otimes_{k=1}^4\mathcal{H}_k$ can be expressed in the Bloch form similar to the
tripartite case \eqref{e:tribloch}. To streamline the notation, all vectors in the Bloch form of $\rho^{ABCD}$
will be denoted as
$R^{A}, R^{AB}, R^{ABC}$ or $R^{ABCD}$ etc. For example, the previous vectors $\bm{a}, \bm{b}, \bm{c}$ etc will be denoted as $R^{A}$, $R^{B}$, $R^{C}$ etc. To express our results, we now rearrange the Bloch vectors $R^{A}, \ldots, R^{ABCD}$ in a matrix form.
As tensor functions, $\rho\longrightarrow R^{i_1\cdots i_s}(\rho)$ is convex linear, i.e. $R^{i_1\cdots i_s}((c\rho_1+(1-c)\rho_2)=cR^{i_1\cdots i_s}(\rho_1)+(1-c)R^{i_1\cdots i_s}(\rho_2)$. 

The realignments are in one-to-one correspondence to biparitions of the index set $\{1 2 3 4\}$ or $\{A, B, C, D\}$.
If the $k$th particle is grouped with $l$th particle, we use underlined indices to indicate such a realignment .
For instance, when the 1st and 3rd particles are grouped together, the column vector $R^{ABCD}$
is converted to a square matrix via 
\begin{equation}
R_{\underline{i}j\underline{l}m}^{\underline{A}B\underline{C}D}=\sum\limits_{i,j,l,m=1}^{d^2-1}R_{ijlm}^{ABCD}\ket{il}\bra{jm},
\end{equation}
where ${\ket{il}}$ (resp. $\bra{jm}$) represents column indices (resp. row indices) in lexicographical order. We will take the freedom to use the same notation for the matrix
$R_{\underline{i}j\underline{l}m}^{\underline{A}B\underline{C}D}=(R_{ijlm}^{ABCD})$ as well.
 Recall that the Ky Fan $k$-norm $\|S\|_k$ of an $m\times n$ matrix $S$ is defined as
 the sum of the $k$th partial sum of the singular values, i.e.
 $\|S\|_k=\sum_{i=1}^{k}\alpha_i$, where $\alpha_i (1\leq \ldots\leq \min(m,n))$ are the singular values of $S$ in decreasing order.

  If a pure state $\ket\Psi\in\otimes_{k=1}^4\mathcal{H}_k $ can be decomposed as $\ket\Psi=\ket\Psi_1\otimes\ket\Psi_2\otimes\ket\Psi_3\otimes\ket\Psi_4$, where $\ket\Psi_k$ is a pure state in $k$th subsystem, then
 $\ket\Psi$ is called fully separable. A pure state $\ket\Psi$ is biseparable provided that it can be written as $\ket\Psi=\ket\Psi_T\otimes\ket\Psi_{\hat{T}}$, where $T$ denotes some subset of subsystems and its complement is $\hat{T}$.
 We now derive some useful bounds for the tensor $R^{ABCD}$, which will be used in detecting multipartite
 entanglement.

Note that there are seven matrix forms (or realignments) of the tensor $R^{ABCD}$, i.e. 7 partitions into two subsets of two particles
 or partitions into one vs. three particles. Namely the $2\times 2$-matrices $\{R_{\underline{i}\underline{j}lm}^{\underline{A}\underline{B}CD}, R_{\underline{i}j\underline{l}m}^{\underline{A}B\underline{C}D}, R_{\underline{i}jl\underline{m}}^{\underline{A}BC\underline{D}}\}$,
 and the $1\times 3$ or $3\times 1$-rectangular matrices $\{R_{\underline{i}jlm}^{\underline{A}BCD}, R_{i\underline{j}lm}^{A\underline{B}CD}, R_{ij\underline{l}m}^{AB\underline{C}D}, R_{ijl\underline{m}}^{ABC\underline{D}}\}$.

We first present upper bounds for the matrix form $R_{\underline{i}\underline{j}lm}^{\underline{A}\underline{B}CD}=\sum\limits_{i,j,l,m}R_{ijlm}^{ABCD}\ket{ij}\bra{lm}$ of the tensor $R^{ABCD}$ in the pure biseparable four-qudit state. The bounds for the other two particles vs two particles matrix forms follow easily by using similar discussion. Moreover, if $\rho^{ABCD}=\rho^{AB}\otimes \rho^{CD}$, then $\mathrm{Tr}(\rho^{ABCD}\lambda_i\otimes\lambda_j\otimes\lambda_l\otimes\lambda_m)=\mathrm{Tr}(\rho^{AB}\lambda_i\otimes\lambda_j)
\mathrm{Tr}(\rho^{CD}\lambda_l\otimes\lambda_m)$, i.e., $R_{ijlm}^{ABCD}=R_{ij}^{AB} R_{lm}^{CD}$.

\begin{lemma} Assume that a pure four-qudit state is biseparable. Then one has that 

1) If the state is fully separable, i.e., for partition $\underline{i}|\underline{j}|l|m$
\begin{equation}
\|R_{\underline{i}\underline{j}lm}^{\underline{A}\underline{B}CD}\|_k\leqslant(d-1)^2;
\end{equation}

2) If the state is biseparable as one vs. three particles, 

(i) for the bipartite partition $\underline{i}|\underline{j}lm$ $(\underline{j}|\underline{i}lm, \text{similarly})$
\begin{equation}\label{1sep3}
\|R_{\underline{i}\underline{j}lm}^{\underline{A}\underline{B}CD}\|_k\leqslant \sqrt k(d-1)\sqrt{(d-1)(d+2)};
\end{equation}

(ii) for the bipartite partition $\underline{i}\underline{j}l|m$ $(\underline{i}\underline{j}m|l, \text{similarly})$
\begin{equation}\label{3sep1}
\|R_{\underline{i}\underline{j}lm}^{\underline{A}\underline{B}CD}\|_k\leqslant \sqrt k(d-1)\sqrt{(d-1)(d+2)};
\end{equation}

3) If the state is separable into two subsystems of two particles, 

(i) for the bipartite partition $\underline{i}\underline{j}|lm$
\begin{equation}\label{2sep2}
\|R_{\underline{i}\underline{j}lm}^{\underline{A}\underline{B}CD}\|_k\leqslant d^2-1;
\end{equation}

(ii) for the bipartite partition $\underline{i}l|\underline{j}m$ $(\underline{i}m|\underline{j}l, \text{similarly})$
\begin{equation}\label{22sep}
\|R_{\underline{i}\underline{j}lm}^{\underline{A}\underline{B}CD}\|_k\leqslant \sqrt k(d^2-1);
\end{equation}
\end{lemma}

\emph{Proof}. We have already shown that $\|R^{A}\|\leqslant\sqrt{d-1}$, $\|R^{AB}\|\leqslant\sqrt{d^2-1}$ (cf.\eqref{Sisobound})
 and $\|R^{ABC}\|\leqslant (d-1)\sqrt{d+2}$ (cf. \eqref{e:sum1}). Also for any matrix $S$, one has that
 $\|S\|_k\leqslant\sqrt k \|S\|$.

1) Therefore for partition $\underline{i}|\underline{j}|l|m$
\begin{align}
\|R_{\underline{i}\underline{j}lm}^{\underline{A}\underline{B}CD}\|_k&=\|R_{\underline{i}}^{\underline{A}}\otimes R_{\underline{j}}^{\underline{B}}\cdot(R_{l}^C\otimes R_{m}^D)\|_k\\\notag
&=\|R_{\underline{i}}^{\underline{A}}\|\|R_{\underline{j}}^{\underline{B}}\|\|R_{l}^C\|\|R_{m}^D\|\leqslant (d-1)^2.
\end{align}

2) (i) As for bipartite partition $\underline{i}|\underline{j}lm$
\begin{align}
\|R_{\underline{i}\underline{j}lm}^{\underline{A}\underline{B}CD}\|_k=\|R_{\underline{i}}^{\underline{A}}\otimes R_{\underline{j}lm}^{\underline{B}CD}\|_k&\leqslant \sqrt k \|R_{\underline{i}}^{\underline{A}}\|\|R_{\underline{j}lm}^{\underline{B}CD}\|\notag\\
&\leqslant\sqrt k(d-1)\sqrt{(d-1)(d+2)},
\end{align}

and similarly one can see it for $\underline{j}|\underline{i}lm$.

\medskip

(ii) For bipartite partition $\underline{i}\underline{j}l|m$ we have that
\begin{align}
\|R_{\underline{i}\underline{j}lm}^{\underline{A}\underline{B}CD}\|_k=\|R_{\underline{i}jl}^{\underline{AB}C}\otimes R_{m}^D\|_k&\leqslant\sqrt k \|R_{\underline{i}jl}^{\underline{AB}C}\|\|R_{m}^D\|\notag\\
&\leqslant \sqrt k (d-1)\sqrt{(d-1)(d+2)},
\end{align}

and the same holds for $\underline{i}\underline{j}m|l$. 

\medskip

3) (i) Now for bipartite partition $\underline{i}\underline{j}|lm$
\begin{align}
\|R_{\underline{i}\underline{j}lm}^{\underline{A}\underline{B}CD}\|_k
&=\|R_{\underline{i}\underline{j}}^{\underline{AB}}\cdot R_{lm}^{CD}\|_k\notag\\
&=\|R_{\underline{i}\underline{j}}^{\underline{AB}}\|\|R_{lm}^{CD}\|\leqslant d^2-1.
\end{align}

(ii) for bipartite partition $\underline{i}l|\underline{j}m$
\begin{align}
\|R_{\underline{i}\underline{j}lm}^{\underline{A}\underline{B}CD}\|_k
=\|R_{\underline{i}l}^{\underline{A}C}\otimes R_{\underline{j}m}^{\underline{B}D}\|_k
&\leqslant\sqrt k\|R_{\underline{i}l}^{\underline{A}C}\otimes R_{\underline{j}m}^{\underline{B}D}\|\notag\\
&=\sqrt k\|R_{\underline{i}l}^{\underline{A}C}\|\|R_{\underline{j}m}^{\underline{B}D}\|\\\notag
&\leqslant \sqrt k (d^2-1),
\end{align}

For $d=2$, one has that
\begin{align}\label{e:bound}
\|R_{\underline{i}\underline{j}lm}^{\underline{A}\underline{B}CD}\|_k&
=\|R_{\underline{i}l}^{\underline{A}C}\otimes R_{\underline{j}m}^{\underline{B}D}\|_k\notag\\
&\leqslant k
\end{align}
due to the fact that the singular values of $A\otimes B$ are products of those of $A$ and $B$.

The same inequality holds for $\underline{i}m|\underline{j}l$ similarly.\qed

\

One can see that $\|R_{\underline{i}\underline{j}lm}^{\underline{A}\underline{B}CD}\|_k\leqslant (d-1)\sqrt{d^2-1}$ for tripartite partition $\underline{i}|\underline{j}|lm$,  $\|R_{\underline{i}\underline{j}lm}^{\underline{A}\underline{B}CD}\|_k\leqslant \sqrt k(d-1)\sqrt{d^2-1}$ for partition $\underline{i}|l|\underline{j}m$ and $\underline{i}|m|\underline{j}l$. It is clear that these two bounds are strictly weaker than the upper bounds in (\ref{1sep3}-\ref{22sep}), which means that if a pure state is tripartite separably it must be biseparable.

Note that if one considers the bipartition of one particle vs three particles, the matrix 
  satisfies that $\|R_{\underline{i}jlm}^{\underline{A}BCD}\|_k\leqslant\sqrt k(d^2-1)$, which is weaker than (\ref{2sep2}).
 Thus we do not take these
 matrix forms into account.

 To detect genuine multipartite entanglement, we define the average matrix $k$-norm of all two vs two partitions 
 $\|M_{22}(R^{ABCD})\|_k=(\|R_{\underline{i}\underline{j}lm}^{\underline{A}\underline{B}CD}\|_k
+\|R_{\underline{i}j\underline{l}m}^{\underline{A}B\underline{C}D}\|_k
+\|R_{\underline{i}jl\underline{m}}^{\underline{A}BC\underline{D}}\|_k)/3$. The following theorem
gives a lower bound for this average norm.

\begin{theorem}\label{thm:GME}
Let $\rho$ be a four-qudit quantum state. If the average $k$-norm satisfies the inequality
\begin{align}\label{t:4}
\|M_{22}(R^{ABCD})\|_k>\max\{\frac{d^2-1+2\sqrt k(d^2-1)}{3}, \sqrt k (d-1)\sqrt{(d-1)(d+2)}\}
\end{align}
for some integer $k\in [1, \ldots, (d^2-1)^2]$, then $\rho$ is genuinely multipartite entangled.
\end{theorem}

\emph{Proof}. Assume that $\rho$ is bipartite separable along the bipartite partition $i|jlm$, then for each $k$
\begin{align}
\|M_{22}(R^{ABCD})\|_k&=\frac13(\|R_{\underline{i}\underline{j}lm}^{\underline{A}\underline{B}CD}\|_k
+\|R_{\underline{i}j\underline{l}m}^{\underline{A}B\underline{C}D}\|_k+\|R_{\underline{i}jl\underline{m}}^{\underline{A}BC\underline{D}}\|_k)\notag\\
&=\frac13(\|\sum_\alpha p_\alpha R_{\underline{i}\underline{j}lm}^{\underline{A}\underline{B}CD}(\rho_\alpha)\|_k+\|\sum_\alpha p_\alpha R_{\underline{i}j\underline{l}m}^{\underline{A}B\underline{C}D}(\rho_\alpha)\|_k\notag\\
&\quad\quad+\|\sum_\alpha p_\alpha R_{\underline{i}jl\underline{m}}^{\underline{A}BC\underline{D}}(\rho_\alpha)\|_k)\notag\\
&\leqslant\frac13\sum_\alpha p_\alpha(\|R_{\underline{i}\underline{j}lm}^{\underline{A}\underline{B}CD}(\rho_\alpha)\|_k
+\|R_{\underline{i}j\underline{l}m}^{\underline{A}B\underline{C}D}(\rho_\alpha)\|_k
+\|R_{\underline{i}jl\underline{m}}^{\underline{A}BC\underline{D}}(\rho_\alpha)\|_k)\\\notag
&\leqslant\sum_\alpha p_\alpha(\sqrt k(d-1)\sqrt{(d-1)(d+2)})\\\notag
&=\sqrt k(d-1)\sqrt{(d-1)(d+2)},
\end{align}
where the second inequality uses (\ref{1sep3}) and (\ref{3sep1}).

If the biseparable is along the bipartite partition $ij|lm$, we can use the bounds (\ref{2sep2}) and (\ref{22sep}) to derive that
$\|M_{22}(R^{ABCD})\|_k\leqslant\frac{d^2-1+2\sqrt k(d^2-1)}{3}$.
Thus, if $\|M_{22}(R^{ABCD})\|_k>\max\{\frac{d^2-1+2\sqrt k(d^2-1)}{3}, \sqrt k (d-1)\sqrt{(d-1)(d+2)}\}$ for some $k$, the quantum state $\rho$ is genuinely multipartite entangled.\qed

\medskip

Example. Let $\rho=(1-p)\ket{\Psi}\bra{\Psi}+\frac{p}{16}I\in\mathcal{H}_2^{\otimes 4}$, where $\ket{\Psi}=\frac12(\ket{0101}+\ket{0110}+\ket{1001}+\ket{1010})$.
It follows from Theorem \ref{thm:GME} that the quantum state $\rho$ is genuinely entangled 
for the white noise tolerance of $p<\frac29$.

\medskip

Remark. The distribution of spin correlations strengths among different subsystems can lead to many interesting
applications. For example, Wang et al\cite{WQ} gave an upper bound for the sum of tripartite spin correlation strength, and obtained a trade-off relation of the Svetlichny inequality for any
multipartite qubits systems by using the upper bound. From Theorem \ref{t:1}, we have $s_{iso}^{AB}+s_{iso}^{AC}+s_{iso}^{BC}\leqslant 1$ for three-qubit state $\rho_{ABC}$. Cheng et al have derived the steering ellipsoid volumes monogamy relation ${v_{B|A}}^{2/3}+{v_{C|A}}^{2/3}\leqslant 1$\cite{CM}. From Corollary \ref{C:2}, we have $s_{iso}^{AB}+s_{iso}^{AC}+s_{iso}^{AD}=\frac{3+\bm a^2-\|R^{BCD}\|^2}{3}$, while it is known that
the monogamy relation ${v_{B|A}}^{2/3}+{v_{C|A}}^{2/3}+{v_{D|A}}^{2/3}\leqslant 1$ holds for pure four-qubit state $\rho_{ABCD}$ by using the conjecture $s_{iso}^{AB}+s_{iso}^{AC}+s_{iso}^{AD}\leqslant1$ (cf \cite{CM}), and the validity of this volume monogamy relation holds for all four qubit state by numerical simulation.

\section{\label{sec:leve5} Summary and discussion}

Spin correlation strengths reveal intrinsic property of bipartite qubits. In this paper, we have generalized the isotropic strength from two-qubit states to three- and four-qudit systems and show that they are also useful concepts for multipartite states and can help analyze
quantum correlations.
In particular, for the tripartite and quadripartite qudit systems, we have obtained the trade-off and various internal bounding relations
of the spin correlations strengths among different subsystems of a multipartite state. We have employed distributions of spin correlations strengths to investigate
quantum entanglement for four-partite qudit systems. We also obtained a criterion to detect genuine multipartite entanglement for any four-qudit state, which generalizes  Vicente-Huber's result for the four-qubit state.

\section*{Acknowledgments}
We thank Jun Li for helpful discussions on entanglement detection and related problems. This work is partially supported by National Natural Science Foundation grant no. 11531004, Simons Foundation grant no. 523868 and a grant from China Scholarship Council.

\bibliographystyle{amsalpha}

\begin{thebibliography}{99}
\bibitem{G} O. Gamel, Phys. Rev. A 93, 062320 (2016).
\bibitem{H09} R. Horodecki, P. Horodecki, M. Horodecki, and K. Horodecki, Rev. Mod. Phys. 81, 865 (2009).
\bibitem{BCP} N. Brunner, D. Cavalcanti, S. Pironio, V. Scarani, and S. Wehner, Rev. Mod. Phys. 86, 419 (2014).
\bibitem{MBC} K. Modi, A. Brodutch, H. Cable, T. Paterek, and V. Vedral, Rev. Mod. Phys. 84, 1655 (2012).
\bibitem{NC} M. A. Nielsen and I. L. Chuang, Quantum computation and quantum information (Cambridge University Press,
Cambridge, 2010).
\bibitem{CKW} V. Coffman, J. Kundu, and W. K. Wootters, Phys. Rev. A 61,052306 (2000).
\bibitem{BBL} P. Badzikag, C. Brukner, W. Laskowski, T. Paterek, and M. Zukowski, Phys. Rev. Lett. 100, 140403 (2008).
\bibitem{V}  J. I. de Vicente, Quantum Inf. Comput. 7, 624 (2007).
\bibitem{HJ} A.S.M. Hassan, P.S. Joag,  Quantum Inf. Comput. 8, 773 (2007)
\bibitem{GT} O. G\"{u}hne and G. T\'{o}th, Phys. Rep. 474, 1 (2009)
\bibitem{VH} J. I. de Vicente and M. Huber, Phys. Rev. A 84, 062306 (2011).
\bibitem{MC} Z. H. Ma, Z. H. Chen, J. L. Chen, Ch. Spengler, A. Gabriel and M. Huber,  Phys. Rev. A 83, 062325 (2011).
\bibitem{CMC} Z. H. Chen, Z. H. Ma, J. L. Chen, and S. Severini, Phys. Rev. A 85, 062320 (2012).
\bibitem{LJ} M. Li, L. Jia, J. Wang, S. Shen, and S. M. Fei, Phys. Rev. A 96, 052314 (2017).
\bibitem{LS} M. Li, S. Shen, N. Jing, S. M. Fei, and X. Q. Li-Jost, Phys. Rev. A 96, 042323 (2017).
\bibitem{OZ} H. Ollivier and W. H. Zurek, Phys. Rev. Lett. 88, 017901 (2001); L. Henderson and V. Vedral, J. Phys. A 34, 6899 (2001).
\bibitem{DVB} B. Daki\'{c}, V. Vedral, and \v{a}. Brukner, Phys. Rev. Lett. 105,190502 (2010).
\bibitem{DVC} W. D\"{u}r, G. Vidal, and J. I. Cirac, Phys. Rev. A 62,062314 (2001).

\bibitem{LF} S. L. Luo and S. S. Fu, Phys. Rev. A 82, 034302 (2010).
\bibitem{WJD} H. M. Wiseman, S. J. Jones, and A. C. Doherty, Phys. Rev. Lett. 98, 140402 (2007).
\bibitem{JPJ} S. Jevtic, M. Pusey, D. Jennings, and T. Rudolph, Phys. Rev. Lett. 113, 020402 (2014).
\bibitem{MJJW} A. Milne, S. Jevtic, D. Jennings, H. Wiseman, and T. Rudolph, New J. Phys. 16, 083017 (2014).
\bibitem{JHAZW} S. Jevtic, M. J. W. Hall, M. R. Anderson, M. Zwierz, and H. M.Wiseman, J. Opt. Soc. Am. B 32, A 40 (2015).
\bibitem{CM} S. Cheng, A. Milne, M. J. W. Hall, and H. M. Wiseman, Phys. Rev. A 94, 042105 (2016).
\bibitem{ZCL}C. Zhang, S. Cheng, L. Li, Q. Y. Liang, et al. Phys. Rev. Lett. 122, 070402 (2019).
\bibitem{B} J. S. Bell, Physics 1, 195 (1965).
\bibitem{CHSH} J. F. Clauser, M. A. Horne, A. Shimony, and R. A. Holt, Phys. Rev. Lett. 23, 880 (1969).
\bibitem{HHH} R. Horodecki, P. Horodecki, and M. Horodecki, Phys. Lett. A 200, 340 (1995).
\bibitem{WQ} Z. Wang, J. Qiao, J. Wang, et al. Quantum Inf Process. 17: 220 (2018).
\bibitem{CH} S. Cheng, M. J. W. Hall, Phys. Rev. Lett. 118, 010401 (2017).

\end{thebibliography}

\end{document}